\documentclass[sigconf]{acmart}

\AtBeginDocument{%
  }

\copyrightyear{2027}
\acmYear{2027}
\setcopyright{cc}
\setcctype{by}
\acmConference[WSDM '27]{Proceedings of the Twentieth ACM International Conference on Web Search and Data Mining}{February 15--19, 2027}{Hong Kong, Hong Kong}
\acmBooktitle{Proceedings of the Twentieth ACM International Conference on Web Search and Data Mining (WSDM '27), February 15--19, 2027, Hong Kong, Hong Kong}
\acmDOI{10.1145/3862322.3863851}
\acmISBN{979-8-4007-3117-4/2027/02}

\acmSubmissionID{1763}

\usepackage{balance}
\usepackage{booktabs}
\usepackage{graphicx}
\usepackage{multirow}
\usepackage{enumitem}
\usepackage[table]{xcolor}
\definecolor{rowblue}{RGB}{238,245,254}
\begin{document}

\title{Learning Better Reasoning for Generative Recommendation with Semantic IDs}

\author{Mengdan Zhu}
\authornote{Work done during an internship at Microsoft.}
\affiliation{%
  \institution{Emory University}
  \city{Atlanta}
     \state{GA}
     \country{USA}
}
\email{mengdan.zhu@emory.edu}

\author{Yufan Zhao}
\affiliation{%
  \institution{Microsoft}
  \city{Redmond}
     \state{WA}
     \country{USA}
}
\email{yufzhao@microsoft.com}

\author{Sophie Di}
\affiliation{%
  \institution{Cornell University}
  \city{Ithaca}
     \state{NY}
     \country{USA}
}
\email{szd5@cornell.edu}

\author{Yao Zhao}
\affiliation{%
  \institution{Microsoft}
  \city{Redmond}
     \state{WA}
     \country{USA}
}
\email{yaozhao2@microsoft.com}

\author{Tao Di}
\affiliation{%
  \institution{Microsoft}
  \city{Redmond}
     \state{WA}
     \country{USA}
}
\email{taodi@microsoft.com}

\author{Yulan Yan}
\affiliation{%
  \institution{Microsoft}
  \city{Redmond}
     \state{WA}
     \country{USA}
}
\email{yulanyan@microsoft.com}

\author{Sridhar Iyer}
\affiliation{%
  \institution{Microsoft}
  \city{Redmond}
     \state{WA}
     \country{USA}
}
\email{sridhariyer@microsoft.com}

\author{Liang Zhao}
\affiliation{%
  \institution{Emory University}
  \city{Atlanta}
     \state{GA}
     \country{USA}
}
\email{liang.zhao@emory.edu}

\renewcommand{\shortauthors}{Mengdan Zhu et al.}

\begin{abstract}
  Generative recommendation reformulates item retrieval as sequence generation, allowing a unified model to directly generate the next item from a user's interaction history. Semantic IDs further make this paradigm effective and scalable by representing each item as discrete codes, enabling knowledge sharing among semantically related items. Recent studies introduce explicit reasoning before Semantic-ID generation, helping models summarize user interests and infer possible preference transitions. However, reasoning is not inherently beneficial: Inaccurate or uninformative reasoning may mislead subsequent item generation and ultimately degrade recommendation performance. This raises a central challenge: how can a recommender select and learn effective reasoning traces and progressively evolve toward better reasoning from its own generations? In this work, we propose Evo-Rec\footnote{Our code is 
available at \url{https://github.com/mengdanzhu/evo-rec}.}, a three-stage framework for learning better reasoning and further enhancing it through reinforcement learning. First, we align Semantic IDs with their textual and behavioral contexts, enabling the model to understand and generate item identifiers. Second, we sample multiple candidate reasoning traces and retain those that improve the prediction of the ground-truth item, providing a stronger reasoning initialization through supervised fine-tuning. Third, we further optimize the reasoning policy through reinforcement learning with catalog-constrained item generation and ranking-aware recommendation feedback. Experiments on three Amazon Review benchmarks show that Evo-Rec consistently outperforms discriminative, generative, and reasoning-enhanced recommenders across all evaluation metrics. On the challenging Video Games dataset, Evo-Rec improves Recall@5 from $0.0710$ to $0.0847$ and NDCG@10 from $0.0563$ to $0.0746$ over the strongest baseline, corresponding to relative gains of $19.3\%$ and $32.5\%$, respectively. These results demonstrate the effectiveness of our framework in learning better reasoning for SID-based generative recommendation.
\end{abstract}





\maketitle

\section{Introduction}
Sequential recommendation aims to predict the next item given a user’s chronologically ordered interaction history.  For a decade the dominant formulation has been discriminative: encode the history into a user vector and score it against a table of item embeddings~\cite{hidasi2018recurrent,tang2018personalized,kang2018self}.
Because every item is mutually independent index in this formulation, the embedding table grows with the item catalog, nothing is shared between semantically related items, and cold items are hard to place. 

Generative recommendation provides an alternative formulation by casting recommendation as autoregressive sequence generation~\cite{geng2022recommendation, rajput2023recommender,pmlr-v235-zhai24a}. In particular, Semantic-ID (SID) based methods quantize item semantics into short sequences of discrete codes, allowing the recommender to generate the identifier of the next item token by token~\cite{hou2023learning,rajput2023recommender} so that related items share code prefixes and generalization no longer depends on interaction counts. Subsequent work has sharpened both ends of this pipeline, learning tokenizers that inject collaborative signal into the codes~\cite{wang2024learnable,wang2024eager,xiao2025unger}, adapting large language models to the code vocabulary through SID--language alignment tasks~\cite{zheng2024adapting,hou2026bridging}, and scaling end-to-end generative recommenders in industrial and open settings~\cite{deng2025onerec,kong2025minionerec}.  Once retrieval is posed as sequence generation over a language model, recommendation inherits the machinery of language models, and most notably their capacity to reason before they answer~\cite{wei2022chain}.

This observation has produced a fast-growing line of reasoning-enhanced recommenders.  One branch reasons in a latent space, deepening the computation applied to a history without emitting any text~\cite{tang2026think,zhang2025reinforced}; another reasons explicitly in
natural language, jointly learning to deliberate and to recommend~\cite{you2026r,kong2025think}, optimizing a reasoning model against a
frozen downstream retriever~\cite{lin2025recr1}, or verbalizing user interests
directly over Semantic IDs~\cite{he2026reasoning}. Despite this progress, existing reasoning-enhanced recommenders largely focus on enabling reasoning, while paying less attention to the quality of the reasoning trajectories. This distinction is important because reasoning is not inherently beneficial for recommendation. For the same interaction history, different reasoning traces may summarize different aspects of user preference, infer different transitions, or even lead to conflicting recommendation decisions. A reasoning trace can therefore be semantically plausible while being misleading for next-item prediction. Training indiscriminately on such traces may not only introduce noisy supervision, but also reinforce spurious reasoning patterns that misguide subsequent reinforcement learning. Similar observations in general LLM reasoning have motivated selection, verification, and self-refinement of reasoning trajectories~\cite{wang2022selfconsistency,zelikman2022star, lightman2023verify}, yet how to identify and improve useful reasoning remains underexplored in generative recommendation.

This limitation persists during reinforcement learning. Existing approaches typically optimize reasoning through feedback derived from the final outcome, such as exact SID matching. Such outcome signals verify whether a prediction is correct, but provide only limited information for improving recommendation rank. In a full-catalog ranking problem, for example, reasoning that moves the ground-truth item from rank 100 to rank 2 is substantially more useful than reasoning that leaves it outside the retrieved set, even though a top-1 exact-match reward assigns both the same reward. Moreover, when reasoning and item generation are optimized as a single trajectory, the quality of the reasoning itself becomes entangled with stochasticity in SID generation. Consequently, the central problem remains underexplored: \textbf{how can a generative recommender identify better reasoning trajectories and continuously improve its reasoning policy from its own generation?}

We approach this problem from the perspective that reasoning can progressively self-evolve through feedback from the recommendation distribution it induces. Given a history $\mathbf{h}$ and a candidate reasoning trace $\mathbf{z}$, useful reasoning should increase the likelihood of the ground-truth item relative to predicting from the history alone, and ultimately move that item toward the top of the catalog ranking. This perspective provides two complementary forms of supervision. Before reinforcement learning, multiple candidate reasoning traces can be compared according to their \emph{predictive utility}, allowing the model to learn from stronger reasoning examples rather than arbitrary teacher generations. During reinforcement learning, the model can generate its own reasoning trajectories and receive ranking-aware feedback according to how each trajectory reorganizes the downstream item ranking. Together, these two mechanisms allow recommendation reasoning to progressively evolve from selected supervision toward task-aligned self-improvement.

Based on this insight, we propose \textbf{Evo-Rec}, a three-stage framework for learning and evolving reasoning in Semantic-ID based generative recommendation. In the first stage, we align newly introduced SID tokens with item semantics and user behavior through multi-task SID--language alignment, providing the representation foundation required for reasoning over itemic tokens. In the second stage, we introduce \emph{Best-of-$N$ Rejection Sampling SFT}. For each interaction history, we sample multiple candidate reasoning traces and measure the incremental predictive utility of each trace by the change in the ground-truth SID likelihood relative to history-only prediction. We retain the highest-utility trace only when it provides a positive improvement and use the selected trajectories to warm up the
reasoning policy. Unlike standard reasoning SFT that treats generated rationales as equally useful, this stage directly optimizes \emph{which reasoning trajectories the model learns from}. The third stage enables the reasoning policy to further evolve from its own generations. For each history, Evo-Rec samples multiple reasoning trajectories from the current policy and, conditioned on each trajectory, performs trie-constrained beam search over the item catalog. We use the rank of the ground-truth SID in the resulting beam to construct an NDCG-based reward and optimize the reasoning policy with GRPO. 

Extensive experiments on three Amazon Review benchmarks demonstrate the effectiveness of Evo-Rec. It consistently outperforms discriminative, generative, and reasoning-enhanced recommendation baselines across all datasets and evaluation metrics. The gains are particularly pronounced on the more challenging Games dataset, where Evo-Rec improves Recall@5 from $0.0710$ to $0.0847$ and NDCG@10 from $0.0563$ to $0.0746$ over the strongest baseline, corresponding to relative improvements of $19.3\%$ and $32.5\%$, respectively. Ablation studies further show that the gains arise primarily from ranking reasoning candidates by predictive utility rather than merely rejecting negative candidates, while ranking-aware NDCG feedback substantially outperforms coarse outcome rewards during reinforcement learning.

Our main contributions are summarized as follows:
\begin{itemize}[leftmargin=*]
    \item We identify reasoning quality as a central bottleneck in
    reasoning-enhanced generative recommendation and propose \textbf{Best-of-$N$ Rejection Sampling SFT} in terms of its predictive utility for the ground-truth item. This allows the model to learn from better reasoning supervision rather than treating all generated rationales as equally useful.

    \item We propose \textbf{Evo-Rec}, a progressive reasoning optimization framework that combines predictive-utility selection with ranking-aware reinforcement learning. The former provides a stronger
    reasoning initialization by selecting useful reasoning supervision, while the latter allows the model to further evolve its reasoning policy from self-generated trajectories.

    \item We introduce a \textbf{catalog-constrained, ranking-aware RL objective} that evaluates each reasoning trajectory through the full recommendation ranking it induces and directly optimizes the reasoning tokens. Extensive experiments and analyses across three domains demonstrate consistent improvements over existing methods and reveal how reasoning selection and ranking-aware feedback contribute to better recommendation.
\end{itemize}

\section{Related Work}

\subsection{From Discriminative to Generative Recommendation}
Sequential recommendation predicts the next item from a chronologically ordered interaction history.  Early approaches are discriminative: they compress the history into a single user representation and score it against an item embedding table.  GRU4Rec models session sequences with recurrent networks and a ranking loss tailored to top-$k$ gains~\cite{hidasi2018recurrent}; Caser applies horizontal and vertical convolutional filters over the embedding matrix of recent interactions to capture point-level and union-level patterns~\cite{tang2018personalized}; and SASRec replaces recurrence with self-attention so that the relevance of every past interaction is weighted adaptively~\cite{kang2018self}.  These models are strong and efficient, but they treat items as mutually independent identifiers.  Consequently, the parameter count of the embedding table grows linearly with the catalog, no representation is shared between semantically related items, and items with few or no interactions are difficult to place in the embedding space, which limits generalization to the long tail.

Generative recommendation reformulates the task as sequence generation and removes the dependence on an explicit item embedding table.  HSTU treats user behavior as a sequential transduction problem and demonstrates that generative recommenders follow favorable scaling behavior at trillion-parameter scale~\cite{pmlr-v235-zhai24a}.  TIGER makes the item vocabulary itself semantic: it quantizes content embeddings with RQ-VAE into a SID, a short sequence of discrete codes, and autoregressively generates the SID of the next item, so that semantically similar items share code prefixes and generalization to unseen items becomes possible~\cite{rajput2023recommender}.  Later work targets the two weak points of this pipeline.  On the tokenizer side, LETTER regularizes item tokenization with collaborative signals and a diversity term, addressing the mismatch between a tokenizer trained for content reconstruction and a recommender trained for preference prediction~\cite{wang2024learnable}.  On the model side, LC-Rec adapts a large language model to the SID vocabulary through a series of alignment tasks that integrate collaborative semantics into the code space~\cite{zheng2024adapting}, and complementary studies benchmark how well LLMs serve as semantic encoders for item representation~\cite{hou2026bridging}. Beyond individual components, OneRec~\cite{deng2025onerec} and MiniOneRec~\cite{kong2025minionerec} study end-to-end generative recommenders and the scaling behavior of SID-based generation in industrial and open settings.  

\subsection{Reasoning-Enhanced Recommendation}
Chain-of-thought prompting shows that generating intermediate steps before an answer substantially improves reasoning in large language models~\cite{wei2022chain}, and recommendation research has adopted this idea along a clear trajectory~\cite{zhu2026learning}.  A group reasons in a latent space: ReaRec performs multi-step latent reasoning inside a sequential recommender at inference time, deepening the computation applied to a history without emitting any text~\cite{tang2026think}.  Latent reasoning is efficient~\cite{zhu2026decompose,li2026latent}, but its intermediate states are neither inspectable nor verifiable, so the quality of the reasoning itself cannot be measured or supervised.  A second group therefore reasons explicitly in natural language.  R$^2$ec learns a unified architecture in which reasoning and recommendation are optimized together~\cite{you2026r}, Rec-R1 optimizes an LLM against feedback from a fixed downstream recommender~\cite{lin2025recr1}, and SIDReasoner reasons explicitly over Semantic IDs and shows that reasoning transfers to generative recommendation~\cite{he2026reasoning}. However, these approaches fail to identify which reasoning trajectories are truly beneficial for downstream recommendation. To address this limitation, our Evo-Rec selects high-utility reasoning trajectories for supervised learning and further refines the policy with ranking-aware reinforcement learning, enabling progressively better recommendation reasoning.

\section{Methodology}

\begin{figure*}[t]
  \centering
  \includegraphics[width=\textwidth]{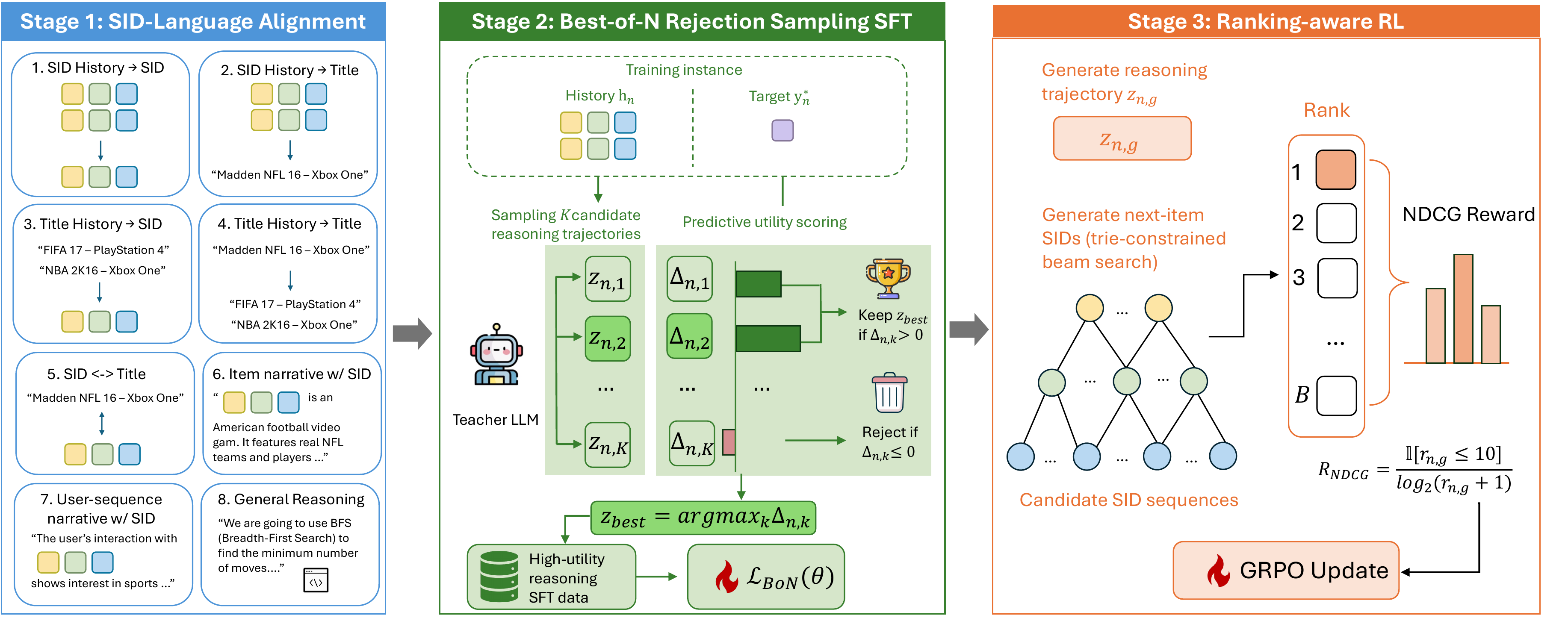}
  \caption{The overview of our proposed \emph{Evo-Rec} framework.}
  \label{fig:main}
\end{figure*}

In this section, we present our \emph{Evo-Rec} framework for progressively learning better reasoning for Semantic-ID based generative recommendation, as illustrated in Figure~\ref{fig:main}. We first formulate reasoning-enhanced next-item generation and describe how Semantic IDs are grounded in language semantics and user behavior. We then introduce Best-of-$N$ Rejection Sampling SFT, which selects reasoning trajectories according to their predictive utility for the target item. Finally, we present ranking-aware reinforcement learning, where self-generated reasoning trajectories are evaluated by the catalog rankings they induce and optimized with NDCG-based feedback, enabling the reasoning policy to progressively evolve toward more effective recommendation reasoning.

\subsection{Problem Formulation}

Let $\mathcal{I}$ denote an item catalog.  A user interaction trajectory is a temporally ordered sequence $S=(i_1,\ldots,i_T,i_{T+1}), i_t\in\mathcal{I}.$
Given the history $H=(i_1,\ldots,i_T)$, the goal of sequential
recommendation is to predict the next item $i_{T+1}$. We cast this task as autoregressive generation by representing every
item with a fixed-length Semantic ID.  Formally, an item tokenizer defines a mapping $\phi:\mathcal{I}\rightarrow\mathcal{C}^{L}, \phi(i)=(c_i^1,\ldots,c_i^L),$
where $\mathcal{C}$ is a discrete code vocabulary and $L$ is the number of code levels. Accordingly, we have the interaction history 
$\mathbf{h}=[\phi(i_1);\ldots;\phi(i_T)],$ and the target SID $\mathbf{y}=\phi(i_{T+1}).$ Instead of directly modeling $p_{\theta}(\mathbf{y}\mid\mathbf{h})$, we introduce an intermediate reasoning sequence
$\mathbf{z}=(z_1,\ldots,z_M)$.  We model the joint
conditional distribution of the reasoning sequence and the target SID
as $p_{\theta}(\mathbf{z},\mathbf{y}\mid\mathbf{h})=p_{\theta}(\mathbf{z}\mid\mathbf{h})p_{\theta}(\mathbf{y}\mid\mathbf{h},\mathbf{z}).$ Thus, we aim to optimize the conditional marginal likelihood of the target SID: $\displaystyle
p_{\theta}(\mathbf{y}\mid\mathbf{h})=\sum\limits_{\mathbf{z}}
p_{\theta}(\mathbf{z}\mid\mathbf{h})
p_{\theta}(\mathbf{y}\mid\mathbf{h},\mathbf{z}).$ 

%Our goal is to learn a better reasoning policy $p_{\theta}(\mathbf{z}\mid\mathbf{h})$ that assigns higher probability to reasoning trajectories that are more useful for downstream recommendation, thereby increasing the likelihood and ranking of the ground-truth item $p_{\theta}(\mathbf{y}\mid\mathbf{h})$.

\subsection{Aligning Language and Semantic IDs in LLMs}

Semantic IDs are newly introduced item tokens and therefore have no inherent meaning to a pretrained language model. Following prior SID--language alignment approaches~\cite{zheng2024adapting,he2026reasoning}, we first align these itemic tokens with natural-language semantics and recommendation behavior before introducing explicit reasoning. For each item $i$, we encode its textual metadata into a continuous representation and apply a residual-quantization VAE (RQ-VAE)~\cite{rajput2023recommender} with $L$ codebooks to obtain $\phi(i)=(c_i^1,\ldots,c_i^L)$. The resulting SID tokens are appended to the LLM vocabulary with randomly initialized embeddings and fine-tune the LLM backbone on a mixture of SID--language tasks following~\cite{he2026reasoning}. This stage provides the semantic and behavioral grounding required by the subsequent reasoning stages.

The first component consists of next-item prediction tasks under different representations of the interaction history and target. Specifically, we include SID history $\rightarrow$ SID, SID history $\rightarrow$ title, title history $\rightarrow$ title, and title history $\rightarrow$ SID. These complementary views expose the model to the same behavioral transition through both natural-language and itemic representations, encouraging knowledge learned in the language space to transfer to the SID space. We further include bidirectional title--SID translation. Beyond these structured prediction tasks, we enrich SID grounding with teacher-generated SID--text interleaved data. At the item level, SID tokens are embedded in natural-language descriptions. At the sequence level, SID interaction histories are interleaved with descriptions of the corresponding user interests and behavioral transitions. These examples expose SID tokens to substantially richer linguistic contexts to help the model associate itemic tokens with semantic concepts useful for subsequent reasoning. In addition, we mix general-domain reasoning examples into the alignment corpus to preserve the pretrained model's general language and reasoning capabilities during recommendation adaptation. All data are optimized under a unified autoregressive objective: 
\begin{equation} 
\mathcal{L}_{\mathrm{align}} = - \mathbb{E}_{\mathbf{x}\sim\mathcal{D}_{\mathrm{align}}} \left[ \sum_{t=1}^{|\mathbf{x}|} m_t(\mathbf{x}) \log p_{\theta}(x_t\mid\mathbf{x}_{<t}) \right], 
\label{eq:alignment} 
\end{equation} 
where $m_t(\mathbf{x})\in\{0,1\}$ specifies whether token $x_t$ contributes to the loss. For recommendation prediction, translation, and general-reasoning examples, we mask the prompt and optimize only the response tokens. For SID--text interleaved examples, we instead apply full-sequence language modeling, i.e., $m_t(\mathbf{x})=1$ for all non-padding tokens, allowing SID embeddings to be learned from every surrounding linguistic context. After this stage, the model is able to interpret and generate SIDs in both semantic and behavioral contexts, providing the representation foundation required for recommendation reasoning. Importantly, Stage 1 does not optimize a particular chain-of-thought trajectory. Rather, it establishes the SID--language grounding on top of which Stage 2 can activate reasoning and learn from selected better reasoning trajectories.

\subsection{Best-of-\texorpdfstring{$N$}{N} Rejection Sampling Supervised Fine-Tuning}

The quality of the CoT reasoning trajectories learned during supervised fine-tuning is critical for subsequent reinforcement learning. The quality of sampled CoTs can vary substantially: a rationale may be plausible in language yet provide little evidence for the correct recommendation.  Directly fine-tuning on such rationales introduces noisy supervision.  We therefore propose Best-of-$N$ Rejection Sampling SFT, which selects CoTs according to their predictive utility for the target SID and rejects candidates that do not improve upon direct
history-based prediction.

\paragraph{Predictive-utility scoring.}
For each training instance $(\mathbf{h}_n,\mathbf{y}_n^*)$, we sample
$K$ explicit CoTs
$\mathcal{Z}_n=\{\mathbf{z}_{n,k}\}_{k=1}^{K}$ from a teacher LLM. We then use a frozen phase-2 model $p_{\bar{\theta}}$ finetuned on a single-sample CoT corpus as a scorer to measure the contribution
of each candidate to target prediction:
\begin{equation}
  \Delta_{n,k}
  = \log p_{\bar{\theta}}
      (\mathbf{y}_n^*\mid\mathbf{h}_n,\mathbf{z}_{n,k})
    - \log p_{\bar{\theta}}
      (\mathbf{y}_n^*\mid\mathbf{h}_n).
  \label{eq:cot-delta}
\end{equation}
Both terms score only the target SID tokens under the same interaction
context and differ solely by the presence of $\mathbf{z}_{n,k}$.
Consequently, $\Delta_{n,k}$ isolates the incremental predictive utility
of the candidate CoT.

\paragraph{Best-of-\texorpdfstring{$N$}{N} selection and rejection.}
We retain the highest-utility candidate only if the predictive utility is positive:
\begin{equation}
  k_n^*=\arg\max_{1\leq k\leq K}\Delta_{n,k},
  \qquad
  \widehat{\mathbf{z}}_n=\mathbf{z}_{n,k_n^*}
  \ \text{if}\ \Delta_{n,k_n^*}>0.
  \label{eq:best-of-n-selection}
\end{equation}
Since the history-only term in Eq.~\eqref{eq:cot-delta} is shared by all candidates, this selection is equivalent to maximizing the
rationale-conditioned likelihood of the target SID.  

\paragraph{Supervised fine-tuning.}
Let $\mathcal{A}=\{n:\Delta_{n,k_n^*}>0\}$ denote the accepted
instances.  We fine-tune the model on the selected reasoning--target
pairs using the joint autoregressive objective:
\begin{equation}
  \mathcal{L}_{\mathrm{BoN}}(\theta)
  =
  -\frac{1}{|\mathcal{A}|}
  \sum_{n\in\mathcal{A}}
  \left[
    \log p_{\theta}
    (\widehat{\mathbf{z}}_n\mid\mathbf{h}_n)
    +
    \log p_{\theta}
    (\mathbf{y}_n^*\mid\mathbf{h}_n,\widehat{\mathbf{z}}_n)
  \right].
  \label{eq:rejection-sft}
\end{equation}
This objective trains the model to generate an outcome-supporting CoT
from the interaction history and subsequently predict the target SID
conditioned on that reasoning.

\subsection{Reasoning Optimization with Constrained SID Ranking}
\label{sec:constrained-rl}

For each history $\mathbf{h}_n$, the current policy samples a group of $G$ reasoning trajectories $\{\mathbf{z}_{n,g}\}_{g=1}^{G}$. Conditioned on each trajectory, we construct a ranked list of valid items with trie-constrained beam search and use the position of the target SID to score the
trajectory.

\paragraph{Constrained beam search.}
Since a valid recommendation must correspond to an item in the
catalog, unconstrained decoding wastes probability mass on invalid SID sequences. We build a prefix trie from all catalog SIDs and restrict each decoding step to valid continuations of the current prefix. Conditioned on a reasoning trajectory $\mathbf{z}_{n,g}$, the score of a complete SID $\mathbf{y}=(y^1,\ldots,y^L)$ is
\begin{equation}
  s_{\theta}(\mathbf{y}\mid\mathbf{h}_n,\mathbf{z}_{n,g})
  =\sum_{\ell=1}^{L}\log \widetilde{p}_{\theta}
  (y^{\ell}\mid\mathbf{h}_n,\mathbf{z}_{n,g},\mathbf{y}_{<\ell}),
  \label{eq:constrained-sid-score}
\end{equation}
where $\widetilde{p}_{\theta}$ is the model distribution after masking and renormalizing over legal trie children. At every level, beam search keeps the $B$ highest-scoring prefixes. The resulting beam $\mathcal{R}_{n,g}=(\widehat{\mathbf{y}}_{n,g}^{(1)},\ldots,
\widehat{\mathbf{y}}_{n,g}^{(B)})$ is an ordered list of distinct and catalog-valid items.

\paragraph{Exact-match and NDCG reward.}
Let $r_{n,g}$ denote the rank of the ground-truth SID $\mathbf{y}_n^*$
in $\mathcal{R}_{n,g}$, with $r_{n,g}=\infty$ if it is absent. The
top-1 exact-match reward and the ranking-aware reward are
\begin{equation}
  R_{n,g}^{\mathrm{EM}}=\mathbb{I}[r_{n,g}=1],
  \qquad
  R_{\mathrm{NDCG}}=\underbrace{\mathbb{I}[r_{n,g}\leq 10]}_{\text{retrieval}} \cdot \underbrace{\frac{1}{\log_2(r_{n,g}+1)}}_{\text{ranking}}.
  \label{eq:ranking-reward}
\end{equation}

% R_{n,g}^{\mathrm{NDCG}@B}
%=\frac{\mathbb{I}[r_{n,g}\leq B]}{\log_2(r_{n,g}+1)}
  
Exact match provides a verifiable but sparse signal: it treats a target ranked second the same as a target absent from the beam. NDCG preserves the exact-match reward of one at rank 1 while assigning progressively smaller credit to lower ranks. We therefore use NDCG@10 as the scalar training reward.

For each history, rewards are normalized across the $G$ trajectories and optimized with GRPO. Consequently, the model learns to favor reasoning paths that rank the ground-truth item more highly.

\section{Experiments}

\subsection{Experimental Setup}

\subsubsection{Datasets and Evaluation.} We evaluate our framework on three public benchmark datasets: Video Games (Games), Office Products (Office), and Industrial and Scientific (Industrial) from the Amazon Review~\cite{hou2026bridging}. We apply 5-core filtering and sort each user's interactions
chronologically.  We truncate each user's historical interaction sequence with a sliding window whose maximum history length is 10.  For each user, we adopt the timestamp split: the earliest 80\% of interactions are used for training, the subsequent 10\% for
validation, and the latest 10\% for testing.  This temporal protocol prevents future interactions from entering the training history.  Dataset statistics
are summarized in Table~\ref{tab:dataset-statistics}.

\begin{table}[t]
  \centering
  \caption{Dataset statistics after preprocessing.}
  \label{tab:dataset-statistics}
  \begin{tabular}{lrrrr}
    \toprule
    Dataset & \#Items & \#Train & \#Val. & \#Test \\
    \midrule
    Games      & 3,858 & 49,133 & 6,142 & 6,142 \\
    Office     & 3,459 & 38,924 & 4,866 & 4,866 \\
    Industrial & 3,686 & 36,259 & 4,533 & 4,533 \\
    \bottomrule
  \end{tabular}
\end{table}

For evaluation, we utilize two widely used metrics: Recall@K (R@K) and NDCG@K (N@K) with cutoff K set to 5 and 10. We follow the full-item ranking setting, where the ground-truth next item is ranked against the entire item catalog rather than a sampled subset of negative items.

\subsubsection{Baselines.} We compare against three families of methods: (1) Traditional discriminative sequential
recommenders include GRU4Rec \cite{hidasi2018recurrent}, Caser \cite{tang2018personalized}, and SASRec \cite{kang2018self}.  (2) Classic generative recommenders include TIGER \cite{rajput2023recommender}, HSTU \cite{pmlr-v235-zhai24a}, LETTER \cite{wang2024learnable}, and LC-Rec \cite{zheng2024adapting}. (3) Reasoning-enhanced recommenders include ReaRec \cite{tang2026think}, R$^2$ec \cite{you2026r}, and SIDReasoner \cite{he2026reasoning}. Baseline results are quoted from \cite{he2026reasoning}, while our method is evaluated using the same data split and evaluation protocol.

\subsubsection{Implementation Details.}

We use Qwen3-1.7B as our LLM backbone and perform full fine-tuning throughout all stages.  Every item is encoded by $L=3$ semantic tokens, and the corresponding SID tokens are appended to the tokenizer vocabulary with randomly initialized embeddings.  Since the three datasets have independent SID codebooks, one model is trained per dataset and no parameter is shared across domains.  All stages are optimized with AdamW on 80GB A100 GPUs. In Stage-1, the backbone is aligned to the SID vocabulary by fine-tuning on a mixture of SID--language tasks.  We use a learning rate of $2\times10^{-5}$, for up to $5$ epochs. For Best-of-$N$ rejection sampling SFT, we synthesize $N=5$ candidate CoTs per training instance by querying the teacher LLM (GPT-5.4) five times independently with the same prompt. The selected instances are then trained for a single epoch with a learning rate of $1\times10^{-5}$ with linear decay and $10$ warm-up steps, a batch size of $72$. The reinforcement learning stage implements GRPO on verl with a vLLM rollout engine over 8*A100 GPUs with batch size of 256.  Each history samples $G=16$ reasoning trajectories.  Conditioned on each trajectory, the SID prefix trie built from the catalog restricts every decoding step, and beam search with $B=10$ returns an ordered list of ten valid items whose target rank defines the NDCG@10 reward of Eq.~\eqref{eq:ranking-reward}. We set the learning rate to $7\times10^{-7}$ and training runs for at most $15$ epochs. At inference time, the model first decodes its reasoning, and the SID is then produced by the same trie-constrained beam search with beam size $10$ over the full item catalog.

\subsection{Main Results}

\begin{table*}[t]
\centering
\caption{The overall performance of different methods on the three datasets. The best results are highlighted in Bold.}
\label{tab:main_results}
\resizebox{\textwidth}{!}{
\begin{tabular}{lcccccccccccc}
\toprule
\multirow{2}{*}{\textbf{Models}}
& \multicolumn{4}{c}{\textbf{Games}}
& \multicolumn{4}{c}{\textbf{Office}}
& \multicolumn{4}{c}{\textbf{Industrial}} \\
\cmidrule(lr){2-5}
\cmidrule(lr){6-9}
\cmidrule(lr){10-13}
& R@5 & N@5 & R@10 & N@10
& R@5 & N@5 & R@10 & N@10
& R@5 & N@5 & R@10 & N@10 \\
\midrule

\multicolumn{13}{c}{\textit{Traditional discriminative sequential recommenders}} \\

\midrule

Caser
& 0.0376 & 0.0241 & 0.0659 & 0.0332
& 0.0880 & 0.0663 & 0.1114 & 0.0738
& 0.0664 & 0.0528 & 0.0852 & 0.0588 \\

GRU4Rec
& 0.0329 & 0.0219 & 0.0599 & 0.0305
& 0.0682 & 0.0480 & 0.0974 & 0.0574
& 0.0788 & 0.0578 & 0.1030 & 0.0649 \\

SASRec
& 0.0501 & 0.0345 & 0.0723 & 0.0416
& 0.1019 & 0.0824 & 0.1167 & 0.0871
& 0.0807 & 0.0647 & 0.0964 & 0.0697 \\

\midrule
\multicolumn{13}{c}{\textit{Classic generative recommenders}} \\
\midrule

TIGER
& 0.0489 & 0.0300 & 0.0763 & 0.0402
& 0.1270 & 0.1037 & 0.1429 & 0.1121
& 0.1003 & 0.0823 & 0.1325 & 0.0924 \\

HSTU
& 0.0539 & 0.0396 & 0.0746 & 0.0462
& 0.1204 & 0.1069 & 0.1323 & 0.1107
& 0.1008 & 0.0898 & 0.1138 & 0.0940 \\

LETTER
& 0.0445 & 0.0294 & 0.0709 & 0.0378
& 0.1315 & 0.1074
& 0.1520 & 0.1139
& 0.1080 & 0.0850
& 0.1389 & 0.0950 \\

LC-Rec
& 0.0441 & 0.0274 & 0.0876 & 0.0412
& 0.0964 & 0.0699 & 0.1487 & 0.0867
& 0.0805 & 0.0520 & 0.1330 & 0.0687 \\

\midrule
\multicolumn{13}{c}{\textit{Reasoning-enhanced recommenders}} \\
\midrule

ReaRec
& 0.0568 & 0.0381 & 0.0843 & 0.0470
& 0.1173 & 0.0988 & 0.1385 & 0.1057
& 0.0973 & 0.0796 & 0.1205 & 0.0870 \\

R$^2$ec
& 0.0655 & 0.0399
& 0.0931 & 0.0525
& 0.1147 & 0.0894 & 0.1486 & 0.1004
& 0.0880 & 0.0774 & 0.1253 & 0.0774 \\

SIDReasoner
& 0.0710 & 0.0460
& 0.1031 & 0.0563
& 0.1373 & 0.1119
& 0.1648 & 0.1208
& 0.1109 & 0.0905
& 0.1438 & 0.1010 \\

\midrule

\rowcolor{rowblue}
\textbf{Evo-Rec}
& \textbf{0.0847} & \textbf{0.0649} 
& \textbf{0.1149} & \textbf{0.0746}
& \textbf{0.1476} & \textbf{0.1223}
& \textbf{0.1710} & \textbf{0.1299}
& \textbf{0.1167} & \textbf{0.0976}
& \textbf{0.1476} & \textbf{0.1076} \\

\bottomrule
\end{tabular}
}
\end{table*}

Table~\ref{tab:main_results} reports the overall performance on the three Amazon Review datasets. Our method consistently outperforms all baselines across every evaluation metric and dataset. These results demonstrate that improving the quality of reasoning trajectories can consistently benefit semantic-ID based generative recommendation across domains. Notably, the improvements are generally more pronounced on NDCG than on Recall. For example, on Games, compared with the strongest baseline, N@5 and N@10 improve by $41.1\%$ and $32.5\%$, while R@5 and R@10 improve by $19.3\%$ and $11.4\%$. This indicates that our method does not merely increase the probability that the target item appears in the retrieved set, but more effectively promotes it toward higher positions in the ranking. This behavior is consistent with our Stage-3 reinforcement learning objective, which directly rewards reasoning trajectories according to the rank of the ground-truth SID under constrained beam search.

The largest gains are observed on Games, where our method raises R@5 from $0.0710$ to $0.0847$ and N@10 from $0.0563$ to $0.0746$. We attribute this improvement to the complementary roles of our two reasoning optimization stages: Stage 2 uses best-of-$N$ selection to provide a stronger reasoning initialization, while Stage 3 further aligns these reasoning trajectories with the downstream ranking objective. Together, the two stages progressively transform higher-quality reasoning supervision into improved recommendation performance.

Games also exhibits substantially lower thinking baseline performance than Office and Industrial in Table~\ref{tab:thinking-stage-comparison}, suggesting that it represents the more challenging recommendation setting and therefore provides greater RL optimization headroom. The particularly large gains on Games suggest that ranking-aware RL can be especially effective when the initial recommendation policy leaves more room for refinement. Moreover, the consistent improvements on Office and Industrial further suggest that the proposed framework is not specific to a single domain, but provides a general mechanism of learning better reasoning for recommendation.

\subsection{Ablation Study}

\begin{table}[t]
  \centering
  \caption{Effect of the CoT selection rule on Games at a fixed sampling budget $N=5$.  All variants share the same candidate CoTs and the same training pipeline, and differ only in which candidate is retained.}
  \label{tab:ablation-selection}
  \resizebox{0.48\textwidth}{!}{
  \begin{tabular}{lccccc}
    \toprule
    Selection rule & R@1 & R@5 & R@10 & N@5 & N@10 \\
    \midrule
    Random selection   & 0.0241 & 0.0694 & 0.1043 & 0.0466 & 0.0587 \\
    Rejection sampling & 0.0248 & 0.0696 & 0.1051 & 0.0489 & 0.0594 \\
    \rowcolor{rowblue}
    \textbf{Best-of-$N$} & \textbf{0.0443} & \textbf{0.0847} & \textbf{0.1149} & \textbf{0.0649} & \textbf{0.0746} \\
    \bottomrule
  \end{tabular}
  }
\end{table}

\begin{table}[t]
  \centering
  \caption{Effect of the sampling budget $N$ on Games under the best-of-$N$ rejection sampling rule.  $N=1$ corresponds to training on a single sampled CoT without selection.}
  \label{tab:ablation-n}
  \begin{tabular}{lccccc}
    \toprule
    Budget & R@1 & R@5 & R@10 & N@5 & N@10 \\
    \midrule
    N=1 & 0.0247 & 0.0690 & 0.1013 & 0.0467 & 0.0571 \\
    N=3 & 0.0277 & 0.0755 & 0.1055 & 0.0518 & 0.0614 \\
    \rowcolor{rowblue}
    \textbf{N=5} & \textbf{0.0443} & \textbf{0.0847} & \textbf{0.1149} & \textbf{0.0649} & \textbf{0.0746} \\
    \bottomrule
  \end{tabular}
\end{table}

\subsubsection{Best-of-$N$ Rejection Sampling.}

\paragraph{CoT Selection Strategy.}

We first isolate the contribution of the CoT selection strategy, which determines \emph{which} reasoning trace is retained. At a
fixed sampling budget of $N=5$, all variants draw the same five candidate CoTs per training instance from the same teacher and differ only in the selection rule: \textbf{random selection} keeps a uniformly drawn from $N$ candidates, \textbf{rejection sampling} keeps the first candidate whose predictive utility is positive
($\Delta_{n,k}>0$), and \textbf{best-of-$N$ rejection sampling} keeps the highest positive utility candidate $\arg\max_k\Delta_{n,k}$ as in Eq.~\eqref{eq:best-of-n-selection}.  Each variant is then trained with the identical Stage-3 recipe, so any difference is attributable to the selection rule alone.

As shown in Table~\ref{tab:ablation-selection} reports the results, Best-of-$N$ rejection sampling improves N@10 by $27.1\%$ over random selection and by $37.1\%$ over rejection sampling, and the gain is most pronounced at the top of the ranking R@1, a relative gain of $83.8\%$.  Two observations follow.  First, random selection pays the full $N$ sampling cost yet performs on par with training on a single trace (Table~\ref{tab:ablation-n}, $N=1$), which shows that the benefit comes from \emph{scoring} the candidates rather than from sampling more of them. Second, rejection sampling is a weak variant: accepting the first positive-utility candidate uses $\Delta_{n,k}$ only as a binary filter and discards the ranking information it carries, and since $94.0\%$ of all candidates are already positive, this filter is close to a random draw.  Ranking candidates by predictive utility is therefore the component that matters.

\paragraph{Sampling Budget.}
We next vary the number of sampled CoTs $N\in\{1,3,5\}$ while keeping the best-of-$N$ rejection sampling rule fixed; $N=1$ degenerates to training on a single sampled CoT without selection.  As shown in Table~\ref{tab:ablation-n}, performance increases monotonically in $N$ on every metric, with N@10 improving from $0.0571$ at $N=1$ to $0.0614$ at $N=3$ and $0.0746$ at $N=5$. These consistent gains demonstrate that our predictive-utility scoring effectively identifies increasingly useful reasoning traces from a larger candidate pool. Moreover, the continued improvement up to $N=5$ suggests that the gains may not have saturated, leaving room for further improvements with larger sampling budgets.

%GPT-5.4 phase1-3

\subsubsection{Sampling Strategy}

We compare the trie-constrained beam search of Section~\ref{sec:constrained-rl},
which returns the $B$ highest-scoring items as an ordered list, against
constrained sampling, which pairs each trajectory with $B$ candidates drawn
independently from the renormalized distribution $\widetilde{p}_{\theta}$ of
Eq.~\eqref{eq:constrained-sid-score}; we set $B=10$ and keep the Stage-2 model the same.  

In this setting, Table~\ref{tab:ablation-sampling} shows that beam search wins on every metric, by $30.9\%$ on N@10 and $56.5\%$ on R@1.  Both decoders share the same trie, so all candidates are catalog-valid either way and the gap is not output validity but of what the reward measures.  Independent draws form an
unordered multiset, leaving the rank $r_{n,g}$ of Eq.~\eqref{eq:ranking-reward} undefined and reducing the reward to a hit indicator, so the policy learns to raise the marginal probability of the target rather than to rank it above its competitors.
Ten unguided draws also cover little of the effective prefixes
of the large Games trie, leaving most groups with zero reward. We therefore adopt constrained beam search as our sampling strategy.

\begin{table}[t]
  \centering
  \caption{Effect of the sampling strategy in RL on Games.  The number of samples and the beam size are both 10.}
  \label{tab:ablation-sampling}
  \resizebox{0.48\textwidth}{!}{
  \begin{tabular}{lccccc}
    \toprule
    Sampling strategy & R@1 & R@5 & R@10 & N@5 & N@10 \\
    \midrule
    Constrained sampling           & 0.0283 & 0.0669 & 0.0943 & 0.0483 & 0.0570 \\
    \rowcolor{rowblue}
    \textbf{Constrained beam search} & \textbf{0.0443} & \textbf{0.0847} & \textbf{0.1149} & \textbf{0.0649} & \textbf{0.0746} \\
    \bottomrule
  \end{tabular}
    }
\end{table}

\subsubsection{Reward Design} 

%prefix/em/ndcg/ndcg+recall

\begin{table}[t]
\centering
\caption{Effect of different reward designs on Games.}
\label{tab:reward-ablation}
\begin{tabular}{lccccc}
\toprule
Reward & R@1 & R@5 & R@10 & N@5 & N@10 \\
\midrule
Prefix Match  & 0.0425 & 0.0791 & 0.1073 & 0.0608 & 0.0699 \\
Exact Match   & 0.0264 & 0.0650 & 0.0958 & 0.0461 & 0.0561 \\
NDCG + Recall & 0.0264 & 0.0633 & 0.0912 & 0.0449 & 0.0538 \\
\rowcolor{rowblue}
\textbf{NDCG Reward}   & \textbf{0.0443} & \textbf{0.0847} & \textbf{0.1149} & \textbf{0.0649} & \textbf{0.0746} \\
\bottomrule
\end{tabular}
\end{table}

We investigate how different rewards affect RL optimization. For each training instance, constrained beam search produces a ranked list of $K=10$ valid SID candidates. We compare four reward designs:

\begin{itemize}[leftmargin=10pt]
    \item Exact Match. We use the exact-match reward defined in Eq.~\ref{eq:ranking-reward}, which gives a positive reward only when the complete target SID is ranked first in the top-10 beam.
    \item Prefix Match. To provide denser supervision, we additionally consider a hierarchical prefix reward. Let $m$ denote the number of consecutive SIDs matching the ground truth and $L$ denote the total number of SIDs. We assign $R_{\mathrm{prefix}}(m)=\frac{1}{2^{L-m}}.$
    \item NDCG Reward. We use the rank-sensitive NDCG reward in Eq.~\ref{eq:ranking-reward}, which assigns a larger reward when the ground-truth SID is ranked higher in the top-$10$ beam.
    \item NDCG + Recall. We also evaluate a hybrid reward that averages the binary Recall@10 signal and the NDCG@10 signal: $R_{\mathrm{NDCG+Recall}} = \left(1+\frac{1}{\log_2(r_{n,g}+1)}\right)/2$, when the ground-truth SID appears in the top-$10$ beam, and $0$ otherwise.
\end{itemize}

As shown in Table~\ref{tab:reward-ablation}, the pure NDCG reward performs best across all evaluation metrics. Exact match performs substantially worse because it treats all top-$10$ hits identically and therefore provides no signal for improving their relative ordering. Adding partial prefix match produces a denser learning signal, but partial SID agreement does not necessarily correspond to ranking the correct item higher. In contrast, the NDCG reward naturally captures both retrieval and ranking: it assigns zero reward when the ground-truth SID is not retrieved in the top-$10$ beam, while discounting successful retrievals according to their rank. Adding an additional Recall@10 term therefore overemphasizes retrieval and weakens the relative ranking signal, which may explain the degraded performance of NDCG+Recall.

%\subsection{Analysis on xxx}

%reasoning length vs performance metrics

\subsection{Discussion}

\subsubsection{Reasoning length}

\begin{figure}[t]
  \centering
  \includegraphics[width=0.9\columnwidth]{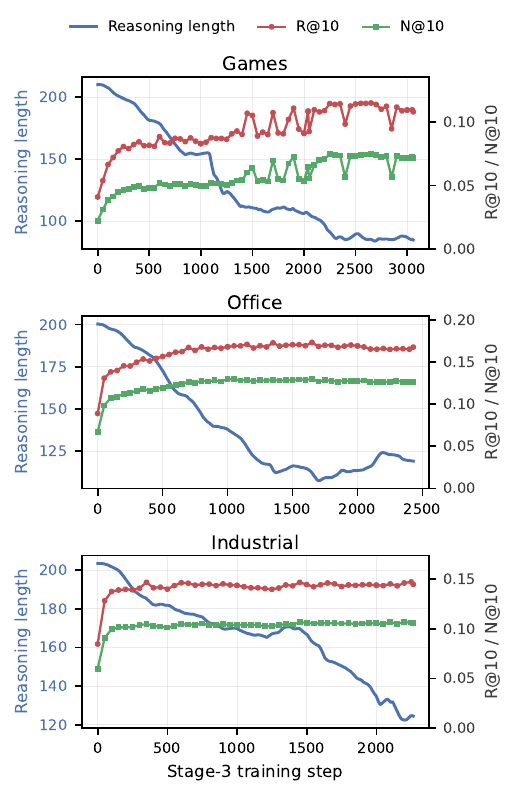}
  \caption{Mean reasoning length over Stage-3 training.}
  \label{fig:reasoning-length}
\end{figure}

Figure~\ref{fig:reasoning-length} shows how reasoning behavior evolves during Stage-3 reinforcement learning. Across all three domains, the average reasoning length decreases substantially during training and eventually converges to a shorter level, while recommendation performance continues to improve. Notably, our objective contains no length penalty; the reward is determined solely by how well a reasoning trajectory ranks the ground-truth SID under constrained beam search. The shortening of reasoning therefore emerges naturally from ranking-aware optimization.

Stage-2 provides the model with better reasoning trajectories as a useful initialization, and during Stage-3 RL, the model progressively learns to disgard those redundant or less relevant reasoning that do not contribute to ranking the target item. This behavior suggests that improved recommendation reasoning does not require increasingly long chains of thought. Instead, effective reasoning appears to depend on how efficiently a trajectory captures information that is useful for ranking. In this sense, Stage-3 RL help the model achieve stronger recommendation performance with shorter and more targeted reasoning trajectories.

%reasoning length vs performance metrics

% \subsubsection{Diversity} compare the model diversity of the first sid. showing the trend of the positive correlation between the diversity and retrieval performance.

\subsubsection{Evolution of Thinking Performance.} 

%think \& non-think comparison, showing the function of different stages 

\begin{table}[t] 
\centering 
\caption{Thinking performance before and after Stage-3 RL.} 
\label{tab:thinking-stage-comparison} 
\resizebox{\columnwidth}{!}{ \begin{tabular}{llccccc} 
\toprule Dataset & Stage & R@1 & R@5 & R@10 & N@5 & N@10 \\ 
\midrule 
\multirow{2}{*}{Games} & Stage 2 & 0.0103 & 0.0256 & 0.0405 & 0.0177 & 0.0225 \\ & Stage 3 & \textbf{0.0443} & \textbf{0.0847} & \textbf{0.1149} & \textbf{0.0649} & \textbf{0.0746} \\  
\midrule
\multirow{2}{*}{Office} & Stage 2 & 0.0495 & 0.0789 & 0.0908 & 0.0652 & 0.0691 \\ & Stage 3 & \textbf{0.0933} & \textbf{0.1476} & \textbf{0.1710} & \textbf{0.1223} & \textbf{0.1299} \\  
\midrule
\multirow{2}{*}{Industrial} & Stage 2 & 0.0435 & 0.0679 & 0.0841 & 0.0562 & 0.0614 \\ & Stage 3 & \textbf{0.0757} & \textbf{0.1167} & \textbf{0.1476} & \textbf{0.0976} & \textbf{0.1076} \\ 
\bottomrule 
\end{tabular} } 

\end{table}

Table~\ref{tab:thinking-stage-comparison} compares recommendation performance under thinking mode before and after Stage-3 reinforcement learning. Although the Stage-2 reasoning policy initially exhibits relatively weak recommendation performance, Stage-3 RL consistently recovers and substantially improves all ranking metrics across the three domains. The improvement is especially pronounced at the top of the ranking. On Games, R@1 increases from $0.0103$ to $0.0446$, corresponding to a $333.0\%$ relative improvement, substantially larger than the gain in R@10. This suggests that Stage-3 optimization does more than recover overall retrieval performance: it sharpens the ranking distribution by pushing the target SID toward the very top of the constrained beam. These results also clarify the different roles of the two training stages. Stage 2 primarily serves as a thinking mode warm-up, teaching the model to produce reasoning in the desired thinking format while using best-of-$N$ rejection sampling to expose it to higher-quality reasoning trajectories. Its objective is therefore not to directly optimize the final ranking performance. Therefore, relatively low thinking performance at this stage does not necessarily imply that the learned reasoning is ineffective. More importantly, learning from better reasoning trajectories provides a stronger potential for subsequent reinforcement learning. Stage-3 RL then provides a task-aligned ranking signal, reinforcing reasoning trajectories that place the ground-truth SID higher in the beam. Consequently, the reasoning policy becomes increasingly aligned with recommendation, yielding large improvements from Stage 2 to Stage 3, particularly at the top of the ranking.

\begin{table}[t]
  \centering
  \caption{Statistics of the best-of-$N$ corpus at $N=5$.  ``pool'' averages all $N$ candidates of an instance, i.e.\ the expected utility of a random pick. }
  \label{tab:bon-corpus}
  \resizebox{0.9\columnwidth}{!}{
  \begin{tabular}{lrrr}
    \toprule
    & Games & Office & Industrial \\
    \midrule
    Instances                          & 49,133  & 38,924  & 36,259 \\
    Candidates ($N{=}5$)               & 245,665 & 194,620 & 181,295 \\
    Saturated candidates               & 1.9\%   & 11.5\%  & 6.7\% \\
    \midrule
    Accepted instances                 & 47,976  & 37,662  & 35,372 \\
    Rejected instances                 & 1,157   & 1,262   & 887 \\
    Rejection rate                    & 2.4\%  & 3.2\%  & 2.4\% \\
    \midrule
    $\bar{\Delta}$ retained           & 3.72    & 3.03    & 4.10 \\
    $\bar{\Delta}$ pool                & 3.16    & 2.60    & 3.61 \\
    Within-instance $\Delta$ range     & 1.22    & 0.95    & 1.07 \\
    \bottomrule
  \end{tabular}
  }
\end{table}

% Single-column variants of both figures are available as
% acceptance_delta_tall and acceptance_slots_tall.
\begin{figure}[t]
  \centering
  \includegraphics[width=0.9\columnwidth]{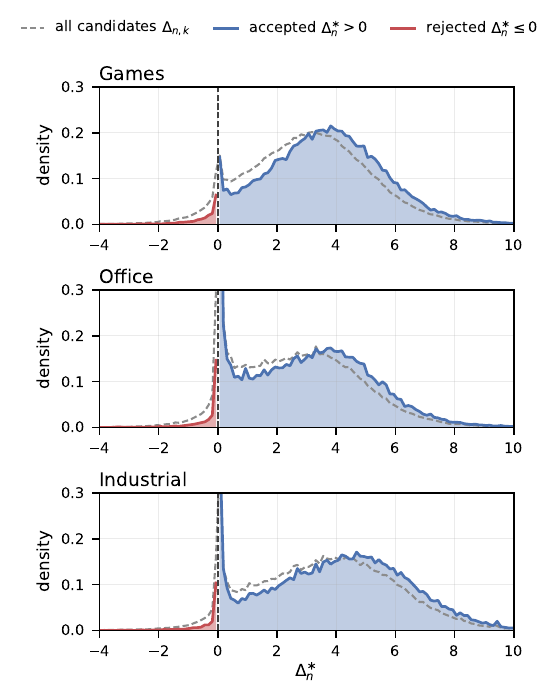}
  \caption{Distribution of the best predictive utility per instance, $\Delta_n^{\ast}=\max_k\Delta_{n,k}$, at $N=5$.}
  \label{fig:acceptance-delta}
\end{figure}
%The clipped spike at $\Delta_n^{\ast}\!\approx\!0$ collects instances whose target is already near-certain from the history alone.
\begin{figure}[t]
  \centering
  \includegraphics[width=0.9\columnwidth]{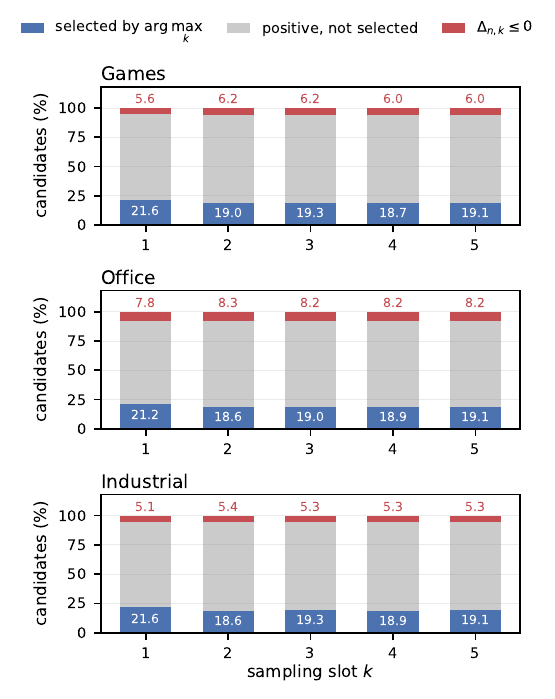}
  \caption{Outcome of the candidates drawn in each sampling slot $k$, as a share of all instances of the domain.}
  \label{fig:acceptance-slots}
\end{figure}

\section{Analysis on Best-of-$N$ Rejection Sampling}

% stats of acceptance n=5 acceptance rate,delta mean,final accepted and rejected data number,delta accepted and rejected distribution,cot accept and reject number for each sampling,

We further analyze the resulting Stage-2 corpus at $N=5$ across all three domains. As shown in Table~\ref{tab:bon-corpus}, the vast majority of sampled reasoning traces have positive predictive utility, and requiring at least one positive candidate rejects only $2.4\%$, $3.2\%$, and $2.4\%$ of training instances. Moreover, rejected instances are predominantly near the decision boundary in  Figure~\ref{fig:acceptance-delta}. These observations explain why rejection sampling alone provides little improvement in Table~\ref{tab:ablation-selection}: the binary $\Delta>0$ criterion is too weak to substantially reshape the training distribution.

Although rejection is rare, the candidate reasoning traces for the same history can differ substantially in predictive utility. The mean within-instance $\Delta$ range, $\max_k \Delta_{n,k}-\min_k\Delta_{n,k}$, is $1.22$, $0.95$, and $1.07$ on Games, Office, and Industrial, respectively. Best-of-$N$ greatly exploits this within-instance variation. Compared with uniformly drawing one candidate from the same pool, the retained $\arg\max_k\Delta_{n,k}$ trace increases predictive utility by $0.56$, $0.43$, and $0.49$ on Games, Office, and Industrial. The main gain comes from ranking candidates within each instance. We found that the dataset with smaller within-instance $\Delta$ range tends to less benefit from Best-of-$N$ rejection sampling. 

Another interesting observation is that a target that is already highly predictable from the history leaves little room for any reasoning trace to further increase its likelihood. Using $\log p_{\bar{\theta}}(y_n^* \mid h_n)>-0.2$ as an indicator of such saturation, we find saturated-candidate rates of $1.9\%$, $11.5\%$, and $6.7\%$ on Games, Office, and Industrial. This effect is particularly pronounced on Office and produces the sharp mass near $\Delta^*\approx0$ in Figure~\ref{fig:acceptance-delta}. Thus, a small $\Delta$ does not necessarily indicate poor reasoning; it can also reflect an instance for which the history-only predictor is already close to its likelihood ceiling. Games exhibits the largest within-instance utility range ($1.22$) and the lowest saturation rate ($1.9\%$), providing the largest opportunity for best-of-$N$ to identify a substantially better reasoning trace. In contrast, Office has both the highest saturation rate ($11.5\%$) and the smallest within-instance range ($0.95$). Its candidate utilities are therefore more tightly compressed, reducing the benefit obtainable from best-of-$N$ selection. 

Taken together, these results reveal that best-of-$N$ rejection sampling operates primarily as a \emph{within-instance ranking mechanism}, with rejection playing a secondary role. Figure~\ref{fig:acceptance-slots} further shows that this gain is not driven by a sampling-slot bias. The five candidate slots are selected by $\arg\max_k \Delta_{n,k}$ at approximately uniform rates across all three datasets, and their negative-utility rates are also largely stable across slots. Thus, no particular sampling position systematically produces better reasoning traces; the benefit arises from comparing multiple candidates and selecting the one with the highest predictive utility. Increasing $N$ therefore provides more opportunities to expose high-utility traces, particularly for instances with large candidate variation. This mechanism is consistent with the monotonic improvement from $N=1$ to $N=5$ in Table~\ref{tab:ablation-n}.

%\section{Transferability/Cross-domain Robustness}

%\section{Case Study}

%accepted/rejected/worse CoT
%phase 2 \& 3

%\enlargethispage{2\baselineskip}

\section{Conclusion}

In this work, we investigate how to learn and progressively improve effective reasoning for Semantic-ID based generative recommendation. We propose \textbf{Evo-Rec}, a three-stage framework that first grounds Semantic IDs in language and user behavior, then selects higher-utility reasoning trajectories
through Best-of-$N$ Rejection Sampling SFT, and finally refines the reasoning policy with catalog-constrained, ranking-aware reinforcement learning. Experiments on three Amazon Review benchmarks demonstrate consistent improvements over discriminative, generative, and reasoning-enhanced recommenders, with particularly strong gains on ranking-sensitive metrics. Our analyses further reveal that the effectiveness of Best-of-$N$ rejection sampling primarily comes from distinguishing reasoning trajectories with different predictive utility within the same interaction history, rather than from rejection alone. Moreover, NDCG-based reinforcement learning provides a more informative signal
than coarse outcome rewards by directly reflecting the ranking induced by each reasoning trajectory. It leads to particularly pronounced gains at the top of the ranking, especially on Recall@1. Overall, these findings support a simple principle: recommendation reasoning should not merely be generated, but should be better selected and optimized according to its downstream recommendation utility.

\section*{Ethical Considerations}
Our experiments are conducted on public Amazon Review benchmarks, which are distributed for research purposes and contain no directly identifying information: users appear as opaque pseudonymous identifiers, and our models take in only interaction order and item-side metadata such as titles and descriptions. Evo-Rec is evaluated in an offline recommendation setting and does not involve deployment on private user data. If applied to real-world systems in the future, appropriate privacy protection, bias auditing, and safeguards against over-personalization should be considered when handling private interaction histories.


\bibliographystyle{ACM-Reference-Format}
\balance
\bibliography{sample-base}

@inproceedings{hou2026bridging,
  title={Bridging language and items for retrieval and recommendation: Benchmarking LLMs as semantic encoders},
  author={Hou, Yupeng and Li, Jiacheng and Fu, Xiangjun and He, Zhankui and Yan, An and Chen, Xiusi and McAuley, Julian},
  booktitle={Proceedings of the 64th Annual Meeting of the Association for Computational Linguistics (Volume 1: Long Papers)},
  pages={3251--3265},
  year={2026}
}

@inproceedings{he2026reasoning,
  title={Reasoning over semantic ids enhances generative recommendation},
  author={He, Yingzhi and Sun, Yan and Tan, Junfei and Chen, Yuxin and Kong, Xiaoyu and Shen, Chunxu and Wang, Xiang and Zhang, An and Chua, Tat-Seng},
  booktitle={Proceedings of the 32nd ACM SIGKDD Conference on Knowledge Discovery and Data Mining V. 2},
  pages={1638--1649},
  year={2026}
}

@article{rajput2023recommender,
  title={Recommender systems with generative retrieval},
  author={Rajput, Shashank and Mehta, Nikhil and Singh, Anima and Hulikal Keshavan, Raghunandan and Vu, Trung and Heldt, Lukasz and Hong, Lichan and Tay, Yi and Tran, Vinh and Samost, Jonah and others},
  journal={Advances in Neural Information Processing Systems},
  volume={36},
  pages={10299--10315},
  year={2023}
}

@InProceedings{pmlr-v235-zhai24a,
  title = 	 {Actions Speak Louder than Words: Trillion-Parameter Sequential Transducers for Generative Recommendations},
  author =       {Zhai, Jiaqi and Liao, Lucy and Liu, Xing and Wang, Yueming and Li, Rui and Cao, Xuan and Gao, Leon and Gong, Zhaojie and Gu, Fangda and He, Jiayuan and Lu, Yinghai and Shi, Yu},
  booktitle = 	 {Proceedings of the 41st International Conference on Machine Learning},
  pages = 	 {58484--58509},
  year = 	 {2024},
  editor = 	 {Salakhutdinov, Ruslan and Kolter, Zico and Heller, Katherine and Weller, Adrian and Oliver, Nuria and Scarlett, Jonathan and Berkenkamp, Felix},
  volume = 	 {235},
  series = 	 {Proceedings of Machine Learning Research},
  month = 	 {21--27 Jul},
  publisher =    {PMLR},
  url = 	 {https://proceedings.mlr.press/v235/zhai24a.html}
}

@inproceedings{zheng2024adapting,
  title={Adapting large language models by integrating collaborative semantics for recommendation},
  author={Zheng, Bowen and Hou, Yupeng and Lu, Hongyu and Chen, Yu and Zhao, Wayne Xin and Chen, Ming and Wen, Ji-Rong},
  booktitle={2024 IEEE 40th International Conference on Data Engineering (ICDE)},
  pages={1435--1448},
  year={2024},
  organization={IEEE}
}

@inproceedings{wang2024learnable,
  title={Learnable item tokenization for generative recommendation},
  author={Wang, Wenjie and Bao, Honghui and Lin, Xinyu and Zhang, Jizhi and Li, Yongqi and Feng, Fuli and Ng, See-Kiong and Chua, Tat-Seng},
  booktitle={Proceedings of the 33rd ACM International Conference on Information and Knowledge Management},
  pages={2400--2409},
  year={2024}
}

@inproceedings{tang2018personalized,
  title={Personalized top-n sequential recommendation via convolutional sequence embedding},
  author={Tang, Jiaxi and Wang, Ke},
  booktitle={Proceedings of the eleventh ACM international conference on web search and data mining},
  pages={565--573},
  year={2018}
}

@inproceedings{kang2018self,
  title={Self-attentive sequential recommendation},
  author={Kang, Wang-Cheng and McAuley, Julian},
  booktitle={2018 IEEE international conference on data mining (ICDM)},
  pages={197--206},
  year={2018},
  organization={IEEE}
}

@inproceedings{hidasi2018recurrent,
  title={Recurrent neural networks with top-k gains for session-based recommendations},
  author={Hidasi, Bal{\'a}zs and Karatzoglou, Alexandros},
  booktitle={Proceedings of the 27th ACM international conference on information and knowledge management},
  pages={843--852},
  year={2018}
}

@article{tang2026think,
  title={Think before recommend: Unleashing the latent reasoning power for sequential recommendation},
  author={Tang, Jiakai and Dai, Sunhao and Shi, Teng and Xu, Jun and Chen, Xu and Chen, Wen and Wu, Jian and Jiang, Yuning},
  journal={IEEE Transactions on Knowledge and Data Engineering},
  year={2026},
  publisher={IEEE}
}

@article{you2026r,
  title={R$^2$ec: Towards Large Recommender Models with Reasoning},
  author={You, Runyang and Li, Yongqi and Lin, Xinyu and Zhang, Xin and Wang, Wenjie and Li, Wenjie and Nie, Liqiang},
  journal={Advances in Neural Information Processing Systems},
  volume={38},
  pages={62376--62405},
  year={2026}
}

@article{deng2025onerec,
  title={Onerec: Unifying retrieve and rank with generative recommender and iterative preference alignment},
  author={Deng, Jiaxin and Wang, Shiyao and Cai, Kuo and Ren, Lejian and Hu, Qigen and Ding, Weifeng and Luo, Qiang and Zhou, Guorui},
  journal={arXiv preprint arXiv:2502.18965},
  year={2025}
}

@article{kong2025minionerec,
  title={Minionerec: An open-source framework for scaling generative recommendation},
  author={Kong, Xiaoyu and Sheng, Leheng and Tan, Junfei and Chen, Yuxin and Wu, Jiancan and Zhang, An and Wang, Xiang and He, Xiangnan},
  journal={arXiv preprint arXiv:2510.24431},
  year={2025}
}

@article{wei2022chain,
  title={Chain-of-thought prompting elicits reasoning in large language models},
  author={Wei, Jason and Wang, Xuezhi and Schuurmans, Dale and Bosma, Maarten and Xia, Fei and Chi, Ed and Le, Quoc V and Zhou, Denny and others},
  journal={Advances in neural information processing systems},
  volume={35},
  pages={24824--24837},
  year={2022}
}

@article{lin2025recr1,
  title={Rec-r1: Bridging generative large language models and user-centric recommendation systems via reinforcement learning},
  author={Lin, Jiacheng and Wang, Tian and Qian, Kun},
  journal={arXiv preprint arXiv:2503.24289},
  year={2025}
}

@inproceedings{zhang2025reinforced,
  title={Reinforced latent reasoning for llm-based recommendation},
  author={Zhang, Yang and Xu, Wenxin and Zhao, Xiaoyan and Wang, Wenjie and Feng, Fuli and He, Xiangnan and Chua, Tat-Seng},
  booktitle={International Conference on Learning Representations},
  volume={2026},
  pages={128449--128470},
  year={2026}
}

@article{kong2025think,
  title={Think before Recommendation: Autonomous Reasoning-enhanced Recommender},
  author={Kong, Xiaoyu and Jiang, Junguang and Liu, Bin and Xu, Ziru and Zhu, Han and Xu, Jian and Zheng, Bo and Wu, Jiancan and Wang, Xiang},
  journal={Advances in Neural Information Processing Systems},
  volume={38},
  pages={141209--141232},
  year={2026}
}

@article{zelikman2022star,
  title={Star: Bootstrapping reasoning with reasoning},
  author={Zelikman, Eric and Wu, Yuhuai and Mu, Jesse and Goodman, Noah},
  journal={Advances in Neural Information Processing Systems},
  volume={35},
  pages={15476--15488},
  year={2022}
}

@inproceedings{geng2022recommendation,
  title={Recommendation as language processing (rlp): A unified pretrain, personalized prompt \& predict paradigm (p5)},
  author={Geng, Shijie and Liu, Shuchang and Fu, Zuohui and Ge, Yingqiang and Zhang, Yongfeng},
  booktitle={Proceedings of the 16th ACM conference on recommender systems},
  pages={299--315},
  year={2022}
}

@inproceedings{hou2023learning,
  title={Learning vector-quantized item representation for transferable sequential recommenders},
  author={Hou, Yupeng and He, Zhankui and McAuley, Julian and Zhao, Wayne Xin},
  booktitle={Proceedings of the ACM Web Conference 2023},
  pages={1162--1171},
  year={2023}
}

@inproceedings{wang2024eager,
  title={Eager: Two-stream generative recommender with behavior-semantic collaboration},
  author={Wang, Ye and Xun, Jiahao and Hong, Minjie and Zhu, Jieming and Jin, Tao and Lin, Wang and Li, Haoyuan and Li, Linjun and Xia, Yan and Zhao, Zhou and others},
  booktitle={Proceedings of the 30th ACM SIGKDD Conference on Knowledge Discovery and Data Mining},
  pages={3245--3254},
  year={2024}
}

@article{xiao2025unger,
  title={Unger: Generative recommendation with a unified code via semantic and collaborative integration},
  author={Xiao, Longtao and Wang, Haozhao and Wang, Cheng and Ji, Linfei and Wang, Yifan and Zhu, Jieming and Dong, Zhenhua and Zhang, Rui and Li, Ruixuan},
  journal={ACM Transactions on Information Systems},
  volume={44},
  number={2},
  pages={1--31},
  year={2025},
  publisher={ACM New York, NY}
}

@article{wang2022selfconsistency,
  title={Self-consistency improves chain of thought reasoning in language models},
  author={Wang, Xuezhi and Wei, Jason and Schuurmans, Dale and Le, Quoc and Chi, Ed and Narang, Sharan and Chowdhery, Aakanksha and Zhou, Denny},
  journal={arXiv preprint arXiv:2203.11171},
  year={2022}
}

@inproceedings{lightman2023verify,
  title={Let's verify step by step},
  author={Lightman, Hunter and Kosaraju, Vineet and Burda, Yuri and Edwards, Harrison and Baker, Bowen and Lee, Teddy and Leike, Jan and Schulman, John and Sutskever, Ilya and Cobbe, Karl},
  booktitle={International Conference on Learning Representations},
  volume={2024},
  pages={39578--39601},
  year={2024}
}

@article{zhu2026learning,
  title={Learning User Interests via Reasoning and Distillation for Cross-Domain News Recommendation},
  author={Zhu, Mengdan and Zhao, Yufan and Di, Tao and Yan, Yulan and Zhao, Liang},
  journal={arXiv preprint arXiv:2602.15005},
  year={2026}
}

@article{zhu2026decompose,
  title={Decompose, Look, and Reason: Reinforced Latent Reasoning for VLMs},
  author={Zhu, Mengdan and Cheng, Senhao and Zhao, Liang},
  journal={arXiv preprint arXiv:2604.07518},
  year={2026}
}

@inproceedings{li2026latent,
  title={Latent implicit visual reasoning},
  author={Li, Kelvin and Shang, Chuyi and Karlinsky, Leonid and Feris, Rogerio and Darrell, Trevor and Herzig, Roei},
  booktitle={Proceedings of the IEEE/CVF Conference on Computer Vision and Pattern Recognition},
  pages={33457--33466},
  year={2026}
}

%\appendix

%\section{Prompts for Corpus Enrichment}

%\section{Baselines}

\end{document}